\documentstyle[epsfig]{elsart}              

\def\gsim{\hbox{\lower3pt\vbox{\baselineskip=4pt \lineskiplimit=0pt \kern2pt
          \hbox{$>$}\hbox{$\sim$}}}}
\def\lsim{\hbox{\lower3pt\vbox{\baselineskip=4pt \lineskiplimit=0pt \kern2pt
          \hbox{$<$}\hbox{$\sim$}}}}
                                                                                
\begin{document}  
\begin{frontmatter}
                                                                              
\title{Short-time behaviour of the two-dimensional \\ 
       hard-disk model} 
\author{A.~Jaster}
\address{Universit\"{a}t - GH Siegen, D-57068 Siegen, Germany}
\date{\today}
                                                                               
%%%%%%%%%%%%%%%%%%%%%%%%%%%%%%%%%%%%%%%%%%%%%%%%%%%%%%%%%%%%%%%%%%%%%%%% 

\maketitle
\begin{abstract}

Starting from the ordered state,
we investigate the short-time behaviour of 
the hard-disk model. For the positional order,
we determine the critical exponents $\eta$ and $z$ from the
dynamic relaxation of the order parameter and the cumulant
with molecular dynamics simulations.
The results are compared with previous Monte Carlo (MC) simulations.
The bond orientational order is studied with MC dynamics.
\end{abstract}

\keyword
{Hard-disk model; short-time dynamics; critical phenomena}
\PACS{64.60.Ht, 82.20.Mj, 05.70.Jk, 64.70.Dv}
\endkeyword

\end{frontmatter}

%%%%%%%%%%%%%%%%%%%%%%%%%%%%%%%%%%%%%%%%%%%%%%%%%%%%%%%%%%%%%%%%%%%%%%%%        
%%%%%%%%%%%%%%%%%%%%%%%%%%%%%%%%%%%%%%%%%%%%%%%%%%%%%%%%%%%%%%%%%%%%%%%%  

\section{Introduction}
For a long time it was believed that universal scaling behaviour
can be found only in the long-time  regime. Therefore, numerical simulations
were performed in the thermodynamic 
equilibrium. Such simulations in vicinity of the critical
point are affected by the critical slowing down. However, recently
Janssen, Schaub and Schmittmann \cite{JASCSC} showed that universality 
exists already in the early time of the 
evolution. They discovered that a system
with non-conserved order parameter and energy (model A) 
quenched from a high temperature 
state to the critical temperature  shows universal short-time behaviour
already after a microscopic time scale $t_{\mathrm{mic}}$.
Starting from an unordered state with a 
small value of the order parameter $m_0$, the order increases 
with a power law 
$M(t) \sim m_0 \, t^\theta$,
where $\theta$ is a new dynamic exponent.
A number of Monte Carlo (MC) investigations \cite{ZHENG,LUSCZH,JAMASCZH} support
this short-time behaviour. These simulations can be also used to calculate the
conventional (static and dynamic) exponents as well as the critical point
\cite{SCHZHE}. 
This may  eliminate critical slowing down, 
since the simulations are performed in the short-time regime.

First simulations of the dynamic relaxation  in the short-time regime
started from an unordered
state. However, short-time dynamical scaling can be also found starting from
the ordered state ($M(t=0)=1$). There exist 
no analytical  calculations for this
situation, but several MC simulations were done
\cite{STAUFFER,LISCZH,ZHENG}. 
Also, all critical exponents, except for the new exponent $\theta$, 
can be calculated starting
from the ordered state. Up to now, simulations of the dynamic relaxation
have only been performed with MC dynamics. 
Normally,  molecular dynamics (MD) simulations can cause ergodicy problems,
since the energy of the system is conserved. However,  for the
two-dimensional hard-disk model the potential
energy of the allowed configurations does not depend on the positions of
the particles, but is constant. Therefore, the restriction of energy conservation does
not lead to a reduction of possible configurations.

The nature of the two-dimensional melting transition is a 
longstanding puzzle \cite{STRAND,GLACLA}. The 
Kosterlitz-Thouless-Halperin-Nelson-Young (KTHNY) theory \cite{KTHNY}
predicts two continuous phase transitions. The first transition
(dislocation unbinding) at the melting 
temperature $T_{\mathrm{m}}$ transforms
the solid  with quasi-long-range positional order and long-range
orientational order into  a  new  {\it hexatic} 
phase that posses short-range positional order and quasi-long-range
orientational order. The  disclination unbinding transition 
at $T_{\mathrm{i}}$ transforms this 
hexatic phase into an isotropic phase in which the  positional and 
orientational order are short-range. There are several other theoretical
scenarios for the melting transition 
in two dimensions \cite{STRAND}. Most of them 
predict a 
first-order phase transition from the solid to the isotropic
phase with a  coexistence region instead of a hexatic phase.

Even for the simple hard-disk system no consensus about the nature of the
melting transition  has been established. 
A large number of simulations of the two-dimensional 
hard-disk model in the thermodynamic equilibrium have been performed.
A melting transition  was 
first seen in a computer study by Alder and Wainwright 
\cite{ALDWAI}. They investigated 870 disks with MD
methods (constant number of particles $N$, volume $V$ and energy $E$)
and found the transition to be first order.
However, the results of such a small system are
affected by large finite-size effects. Recent simulations used 
MC techniques either in the $NVT$ ensemble (constant volume)  
\cite{ZOLCHE,WEMABI,MIWEMA,JASTER} or the  $NpT$ ensemble (constant
pressure) \cite{LEESTR,FEALST}. 
Unfortunately, the results of these simulations are 
not compatible.

In this article, we study the short-time behaviour of
the two-dimensional hard-disk model
starting from the ordered state (perfect crystal). 
First we examine the positional order parameter 
$\psi_{\mathrm{pos}}(t)$ and the cumulant $\tilde{U}_{\mathrm{pos}}(t)$
with MD simulations. The power law behaviour of these observables
is used to determine
the critical exponents $\eta$ and $z$.
The results are compared with those of a previous MC
simulation \cite{JASTER2}. In the second part we study the bond orientational
order parameter $\psi_6(t)$ and the cumulant $\tilde{U}_6(t)$
with MC dynamics. All simulations are performed in
a rectangular box with ratio $2:\sqrt{3}$,
which is necessary for the ordered state, 
and with periodic boundary conditions. 
The disk diameter is set equal to one.

\section{Positional order}
The positional order parameter $\psi_{\mathrm{pos}}$ can be computed via
\begin{equation}
\psi_{\mathrm{pos}}(t) = \left | 
\frac{1}{N} \sum_{i=1}^N \exp ( {\mathrm{i}} \, \vec{G} \cdot \vec{r}_i(t) )
\right | \ ,
\end{equation}
where $\vec{G}$ denotes a reciprocal lattice vector and $\vec{r}_i(t)$ 
is the position of  particle $i$ at time $t$. $\vec{G}$ has a magnitude of 
$2 \pi /a$, where $a=\sqrt{2/(\sqrt{3}\rho)}$ is the average 
lattice spacing. The direction of $\vec{G}$ is fixed to that of  
a reciprocal lattice vector of the perfect crystal (which are unique due
to the boundary condition of a rectangular box of ratio $2:\sqrt{3}$).
The reason for fixing $\vec{G}$
is that large crystal tilting  is not possible 
since we simulate only the short-time behaviour of the system.

MC simulations of the dynamic relaxation 
of systems with quasi-long-range order were performed  for the
6-state clock model \cite{CZERIT}, the XY model \cite{OKSCYAZH,LUOZHE},
the fully frustrated XY model \cite{LUSCZH,LUOZHE}, the quantum XY
model \cite{YILUSCZH} and the 
hard-disk model \cite{JASTER2}.
However, no investigations exist for the relaxation with molecular dynamics.
Independent of the dynamics,
the scaling form of the second moment of the order parameter at or above 
$\rho_{\mathrm{m}}$ is
\begin{equation}
\label{Eqscale}
{\psi_{\mathrm{pos}}}^2(t,L) = b^{-\eta}  
{\psi_{\mathrm{pos}}}^2(b^{-z}t,b^{-1}L) \ , 
\end{equation}
where $b$ denotes the rescaling factor. This
leads  for sufficiently large $L$ to a power law time dependence  of the form
\begin{equation}
\label{EqPLpsi2}
{\psi_{\mathrm{pos}}}^2(t) \sim t^{-\eta/z} \ .
\end{equation}
From a finite-size scaling (FSS) analysis of the
time-dependent cumulant 
\begin{equation}
\tilde{U}_{\mathrm{pos}}(t)= \frac{{\psi_{\mathrm{pos}}}^4(t)}{
\left ( {\psi_{\mathrm{pos}}}^2(t) \right ) ^2} -1
\end{equation}
one obtains
\begin{equation}
\tilde{U}_{\mathrm{pos}}(t) \sim t^{d/z} \ ,
\end{equation}
where $d$ is the dimension of the system.
One can use $\tilde{U}_{\mathrm{pos}}(t)$ to  determine the dynamic critical exponent $z$
and then, with $z$ in hand, the static exponent
$\eta$ from the behaviour of ${\psi_{\mathrm{pos}}}^2(t)$.

In the following we simulate the hard-disk model with molecular dynamics
at the melting density $\rho_{\mathrm{m}} \approx 0.933$ \cite{JASTER2}
(the density is given as usual in reduced units)
and in the solid phase ($\rho=1.0$) to investigate the time evolution
of ${\psi_{\mathrm{pos}}}^2$ and $\tilde{U}_{\mathrm{pos}}$. 
Since the positional order of the system is quasi-long-range,
we expect to find a power law behaviour for $\rho \ge \rho_{\mathrm{m}}$
as in the case of the MC study.
Starting from the ordered
state ($\psi_{\mathrm{pos}}=1$), i.e.\ from a perfect crystal with lattice
spacing $a$, we release the system to evolve with molecular dynamics.
Solving exactly the simultaneous 
classical equations of motion for hard-disks means
in practice to calculate the next collision point of two particles and their 
new momenta. 
The initial components of the momenta are chosen randomly with a distribution 
proportional $\exp (p^2_i/2)$. 
Obviously, in the case of  a hard-core potential  a global change in the kinetic 
energy just leads to a rescaling of time.
We use systems of $16^2$, $32^2$,
$64^2$ and $128^2$ hard-disks and measure the observables up to $t=6$
in time intervals of $\Delta t=0.04$. 
The number of independent data sets ranges from ${\cal O}(150)$ for 
$N=128^2$ to ${\cal O}(30\,000)$ for $N=16^2$. 
Statistical errors are calculated by
dividing the data into different subsamples. Systematic errors are
estimated by least square fits
of different system sizes and different time
intervals, i.e.\ the exponent $c$ of the power law behaviour 
is calculated for each system in small
intervals $[t_i,t_j]$. 
For large enough systems, $c$ is independent of the system
size between $t_{\mathrm{min}}$ and $t_{\mathrm{max}}$.
The microscopic time scale $t_{\mathrm{min}}$ depends on the microscopic
details, but is independent from the system size. In contrast to this,
$t_{\mathrm{max}}$ --- the time when the system starts to show finite-size
effects --- scales with the number of particles.

\begin{figure}[t]
\begin{center}
\mbox{\epsfxsize=12.0cm
\epsfbox{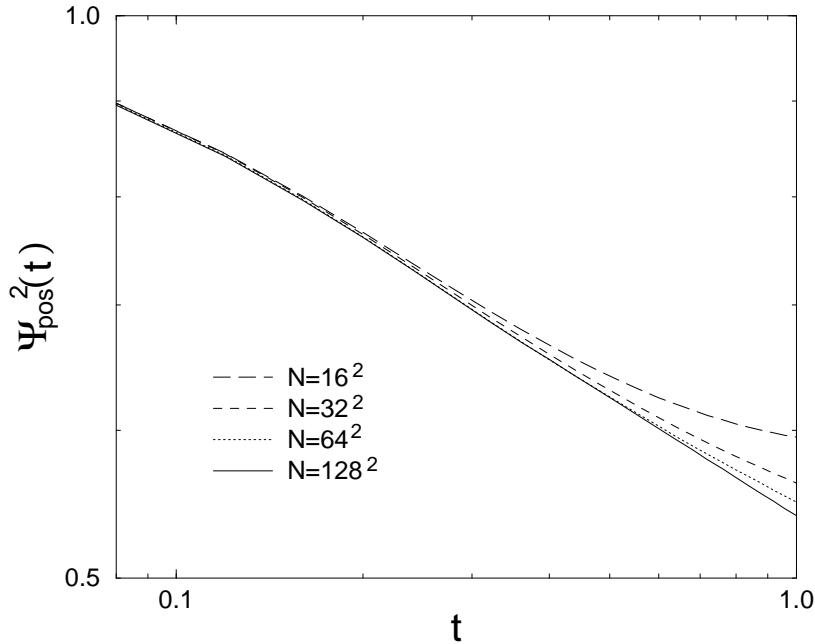}}
\end{center}
\caption{\label{fig_pos_O2}
Second moment of the positional order
parameter ${\psi_{\mathrm{pos}}}^2$ as a function of time
starting from the  ordered state at
$\rho=0.933$. ${\psi_{\mathrm{pos}}}^2(t)$ was calculated in time steps
of $\Delta t=0.04$, where we used MD simulations.}
\end{figure}

In Fig.\ \ref{fig_pos_O2}  we plot the time evolution of 
${\psi_{\mathrm{pos}}}^2$ at the melting density 
$\rho_{\mathrm{m}}$ for different system sizes in
a double logarithmic scale. 
The figure shows that the power law behaviour starts after
a microscopic time scale $t_{\mathrm{mic}}$ of approximately $0.16$. 
For times up to  $0.5$ the difference between the systems with
$64^2$ and $128^2$ hard-disks is negligible. 
We use these two systems and time
intervals of $t=[0.2 \dots 0.8]$ or smaller for the determination
of the critical exponents.
Power law fits for the different 
time intervals and system sizes lead to $\eta/z = 0.201(4)$.
The situation in the solid phase at $\rho = 1.0$ is very similar 
and we get $\eta / z = 0.0695(25)$.
To determine $z$ independently, we also measure the time evolution of the
cumulant $\tilde{U}_{\mathrm{pos}}$. 
As before the behaviour can be well described by a power law ansatz.
From the slope we get $z=1.04(3)$ at $\rho_{\mathrm{m}}$, while the analysis
for $\rho=1.0$ yields $z=1.06(3)$. Thus we get $\eta=0.194(9)$ at
$\rho=0.933$ and $\eta=0.0737(49)$ at $\rho=1.0$, respectively. The values 
of $\eta$ coincide with
the results from the dynamic relaxation
with MC dynamics \cite{JASTER2} (using the same methods) 
within statistical errors, as can be seen from Table
\ref{table1}. For comparison we also show the results
obtained from conventional FSS \cite{JASTER2}. 
The value of the dynamic critical exponent 
$z$ changes from $z \approx 2$ for MC dynamics to $z \approx 1$ 
for MD simulations. These are the usual values for local MC updating 
schemes --- which can be understood as due to the `diffusion' of the
changes induced by the local updating --- and MD simulations, respectively. 
The time when the system 
starts to shows finite-size effects
scales approximately
with $L^z$ for MD simulations (as can be estimated from Fig.\ 
\ref{fig_pos_O2}) as well as for  MC investigations.
%%%%%%%%%%%%%%%%%%%%%%%%%%%%%%%%%%%%%%%%%%%%%%%%%%%%%%%%%%%%%%%%%%%%%%%% 
%%%%%%%%%%%%%%%%%%%%%%%%%%%%%%%%%%%%%%%%%%%%%%%%%%%%%%%%%%%%%%%%%%%%%%%% 
%%%%%%%%%%%%%%%%%%%%%%%%%%%%%%%%%%%%%%%%%%%%%%%%%%%%%%%%%%%%%%%%%%%%%%%%
\begin{table}[b]
\caption{ \label{table1}
The critical exponents $z$ and $\eta$ determined from the 
short-time behaviour of $\tilde{U}_{\mathrm{pos}}(t)$ and 
${\psi_{\mathrm{pos}}}^2(t)$ 
with MD and MC dynamics [16]
and the value of $\eta$ measured with FSS methods [16].}
%\begin{center}
\begin{tabular}{cc*{2}{r@{.}l}c*{2}{r@{.}l}c*{1}{r@{.}l}}
%\vspace*{-4.0mm} \\
\hline
\hline
\multicolumn{1}{c}{ }  & 
\multicolumn{1}{c}{\phantom{xxxx}}  & \multicolumn{4}{c}{MD short-time} &
\multicolumn{1}{c}{\phantom{xx}}  &\multicolumn{4}{c}{MC short-time} &
\multicolumn{1}{c}{\phantom{xx}}   &\multicolumn{2}{c}{FSS} \\ 
$\rho$  & & \multicolumn{2}{c}{$z$} &  
\multicolumn{2}{c}{$\eta$}  &  &
\multicolumn{2}{c}{$z$}  &  
\multicolumn{2}{c}{$\eta$} & & 
\multicolumn{2}{c}{$\eta$} \\
\hline
0.933 & & 1&04(3) & 0&194(9) & &  2&01(2) &  0&199(3)  & &  0&200(2) \\
1.0   & & 1&06(3) & 0&0737(49) & & 2&06(4)  &  0&0794(29) & &  0&0791(6) \\
\hline
\hline
\end{tabular}
%\end{center}
\end{table}
%%%%%%%%%%%%%%%%%%%%%%%%%%%%%%%%%%%%%%%%%%%%%%%%%%%%%%%%%%%%%%%%%%%%%%%%

Our results show that the determination of the critical exponents in the
short-time region can be done with molecular dynamics at least
in the case of hard-core potentials. The disadvantage of the MD
simulations compared to the MC investigations with the Metropolis
algorithm is the scaling of CPU time with the number of particles.
In the Metropolis algorithm the CPU time for a sweep is proportional 
to the number of particles $N$, i.e.\ a single step of a disk 
is independent of $N$. In the MD simulations the number of collisions
during a fixed time interval $\Delta t $ scales also with $N$. However,
the search for the next two colliding disks is not independent of $N$.
The CPU time for
a naive algorithm scales with $N^2$, while an improved version yields
a factor $N$. Therefore, also the improved MD 
algorithm leads to a poorer scaling 
behaviour as the MC version.
However, the MD simulations not only enlarge the knowledge of dynamic 
relaxation but perhaps also offer a possibility to compare the results with
experimental investigations.

\section{Bond orientational order}
The orientational order of the hard-disk system can be described by the bond
orientational order parameter 
\begin{equation}
\psi_6 =  \frac{1}{N} \sum_{i=1}^{N} 
\frac{1}{N_i} \sum_{j=1}^{N_i} \exp ( 6 \, {\mathrm{i}}\, \theta_{ij} )
\ .
\end{equation} 
The sum on $j$ is over the $N_i$ neighbours of the 
particle $i$ and $\theta_{ij}$ is the angle between the particles $i$
and $j$ and an arbitrary but fixed reference axis. Two particles are
defined as neighbours, if the distance is less than $1.4$ times the average 
lattice spacing $a$. This definition is computationally less expensive than
the precise determination with the Voronoi construction \cite{VORONOI}.

We examine the bond orientational order of the hard-disk model in the liquid
regime at $\rho=0.885$, in vicinity of the transition point 
$\rho_{\mathrm{i}} \approx 0.899$ \cite{JASTER} at $\rho=0.9$ and 
$\rho=0.905$ and in the solid phase at
$\rho=0.94$. We measure the second moment of the order parameter
${\psi_6}^2$ and the cumulant $\tilde{U}_6$ as a function of time, where
the MC dynamics is given by the Metropolis algorithm \cite{METRO}. 
The new positions of the particles are chosen with equal probability
within a circle centered about its original position. As before,
we start the relaxation process from the perfect ordered crystal 
($\psi_6=\psi_{\mathrm{pos}}=1$). We use systems of $64^2$ and
$128^2$ hard-disks and measure up to $10\,000$ MC sweeps. 

In case of a KTHNY-like scenario $\rho_{\mathrm{i}}$ is the beginning
of the hexatic phase, i.e., the lower bond of the critical line
which ends at $\rho_{\mathrm{m}}$. Therefore,
we expect a power law behaviour similar to Eq.\ (\ref{EqPLpsi2}) 
between $\rho_{\mathrm{i}}$ and $\rho_{\mathrm{m}}$
assuming a scaling behaviour of the form 
(\ref{Eqscale}) for the positional order parameter.
The value of the critical exponent $\eta_6$ (at $\rho=\rho_{\mathrm{i}}$)
is predicted with $1/4$
\cite{KTHNY} and was measured with 0.251(36) \cite{JASTER}, while
the value of the dynamic critical exponent  $z$
for the local Metropolis algorithm is normally about two.
For a conventional weak first-order phase transition the behaviour of 
the positional order parameter should be also approximately power-like.

%%%%%%%%%%%%%%%%%%%%%%%%%%%%%%%%%%%%%%%%%%%%%%%%%%%%%%%%%%%%%%%%%%%%%%%%
\begin{figure}
\begin{center}
\mbox{\epsfxsize=12.0cm
\epsfbox{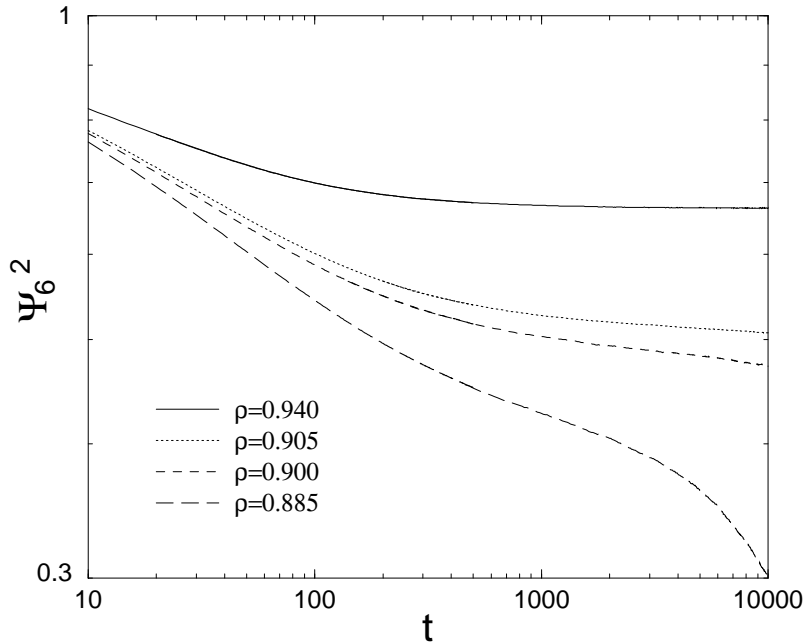}}
\end{center}
\caption{\label{fig_or_O2}
Time evolution of the second moment of the bond orientational order
parameter ${\psi_6}^2$ starting from the  ordered state at
$\rho=0.885$, $0.900$, $0.905$ and $0.940$. The MC dynamics is given by 
the Metropolis algorithm.}
\end{figure}
\begin{figure}
\begin{center}
\mbox{\epsfxsize=12.0cm
\epsfbox{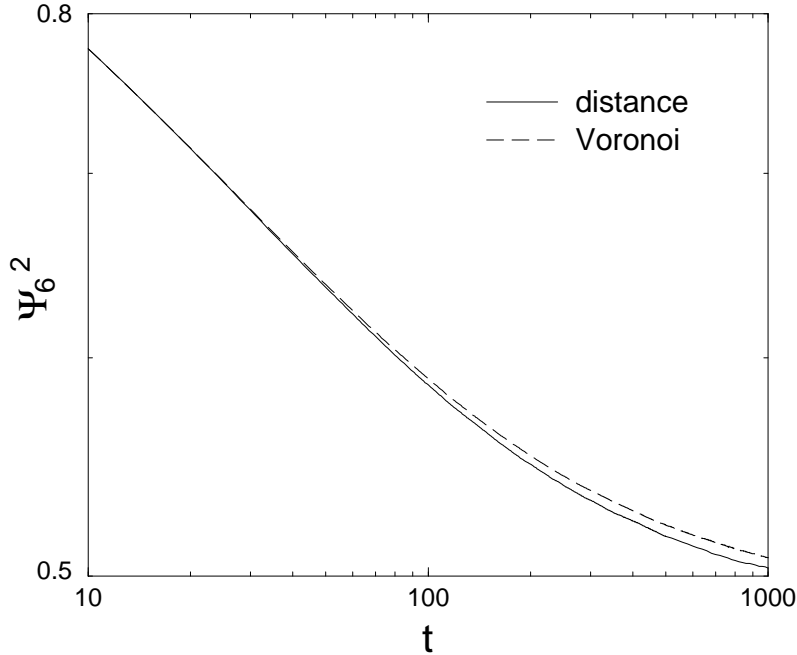}}
\end{center}
\caption{\label{fig_or_voro}
Time evolution of the second moment of the bond orientational order
parameter ${\psi_6}^2$ starting from the  ordered state at
$\rho=0.9$. The solid curve is the same as in Fig.\ 2, i.e.\ neighbours are
defined with the distance of the particles. The dashed line shows the 
behaviour that we obtained using the Voronoi construction for the 
determination of neighbours.}
\end{figure}

%%%%%%%%%%%%%%%%%%%%%%%%%%%%%%%%%%%%%%%%%%%%%%%%%%%%%%%%%%%%%%%%%%%%%%%%
Figure \ref{fig_or_O2} shows the time evolution of ${\psi_6}^2$ for the
different densities. Obviously, the time dependence in vicinity
of $\rho_{\mathrm{i}}$ can not be described by a simple power law
behaviour as expected. Therefore, the scaling form of the second moment
of the order parameter ${\psi_6}^2$ is not given by (\ref{Eqscale}).
The behaviour of ${\psi_6}^2(t)$ at $\rho=0.9$ and $0.905$
is also not consistent with
a simple weak first-order phase transition. The
behaviour in the fluid phase ($\rho = 0.885$) is incompatible 
with both scenarios, since it is not 
of the form  $t^{-\eta_6/z}  \exp (-t/\xi_{\mathrm{t}})$ \cite{JASTER3}.
To rule out that the non power law behaviour is just an effect
coming from our method determing neighbours, we also make 
simulations using the precise Voronoi definition. The result for
${\psi_6}^2(t)$ at $\rho=0.9$ is visualized in Fig.\ \ref{fig_or_voro}.
For small times the difference between both definitions is negligible.
At larger times the number of disclinations increase and causes
errors in case of  using the distance for the determination of neighbours.
Therefore, the deviation between the different values of ${\psi_6}^2$
grows if the time increases. 
However, also in case of using the Voronoi construction
we find no power law dependence.
%%%%%%%%%%%%%%%%%%%%%%%%%%%%%%%%%%%%%%%%%%%%%%%%%%%%%%%%%%%%%%%%%%%%%%%%
%%%%%%%%%%%%%%%%%%%%%%%%%%%%%%%%%%%%%%%%%%%%%%%%%%%%%%%%%%%%%%%%%%%%%%%%

\begin{figure}
\begin{center}
\mbox{\epsfxsize=12.0cm
\epsfbox{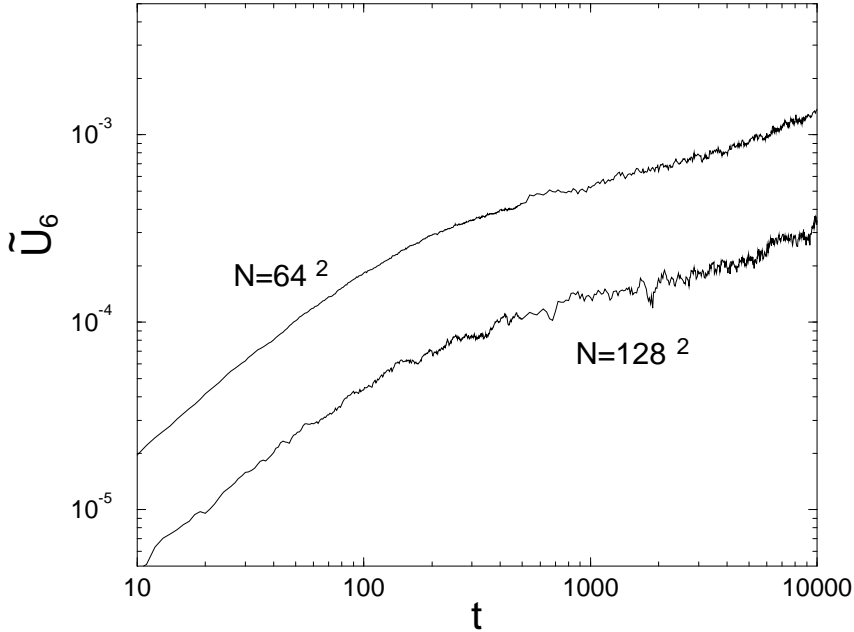}}
\end{center}
\caption{\label{fig_or_Ut}
Time dependent cumulant $\tilde{U}_6(t)$ at $\rho=0.9$ for $N=64^2$
and $N=128^2$ for the Metropolis algorithm.}
\end{figure}

A possible explanation for the behaviour of ${\psi_6}^2(t)$ 
could be a mixing of the order parameters\footnote{A mixing of
order parameters was also found for the Ashkin-Teller model 
\cite{LILIPASC}.}. 
In the simplest case we expect 
that the behaviour should be described by the sum of two power law functions.
However, we find that the
the curves at $\rho=0.9$ and $\rho=0.905$ could not
be fitted very well by such an ansatz. Nevertheless, we
use the almost linear behaviour in the intervals $t_1=[10,100]$
and $t_2=[1000,10\,000]$ to determine the values of the exponents,
i.e.\ we examine if the behaviour in one of these intervals is consistent
with a power law coming from the bond orientational order parameter.
At $\rho=0.9$ we get from ${\psi_6}^2(t)$ the exponent
$c_1=0.124$ and $0.0264$, respectively.
The slope of the cumulant $\tilde{U}_6(t)$ (which is shown in
Fig.\ \ref{fig_or_Ut}) gives the exponent $c_{\mathrm{U}}$. 
This yields $z=2.12$, $\eta=0.262$ (for $t_1=[10,100]$)
and $z=4.95$, $\eta=0.130$ (for $t_2=[1000,10\,000]$). 
The first value of $\eta$
is consistent with previous measurements of $\eta_6$ 
in equilibrium \cite{JASTER},
while the value determined in the interval $t_2$ is too small. However,
also the behaviour of  ${\psi_6}^2(t)$ in the  interval $t_1$ 
is inconsistent with (\ref{Eqscale}) 
since the value is not constant in the solid phase.
Thus, neither the behaviour in the time 
interval $t_1$  nor in the interval $t_2$ is compatible with
the scaling form (\ref{Eqscale}).
Additionally, we try to estimate the dynamic critical exponent $z$
from the exponential behaviour the the fluid phase. But these
measurements give also inconsistent results, i.e.\ we get a value 
$z\approx 1$. 
%%%%%%%%%%%%%%%%%%%%%%%%%%%%%%%%%%%%%%%%%%%%%%%%%%%%%%%%%%%%%%%%%%%%%%%%
%%%%%%%%%%%%%%%%%%%%%%%%%%%%%%%%%%%%%%%%%%%%%%%%%%%%%%%%%%%%%%%%%%%%%%%%

\section{Conclusions}
We investigated the short-time behaviour of the hard-disk model for
the positional and bond orientational order. The positional order parameter
was studied with MD simulations at $\rho_{\mathrm{m}}$ and in the 
solid phase. We determined the critical exponent $\eta$ as well
as the dynamic critical exponent $z$ from the power law behaviour of
${\psi_{\mathrm{pos}}}^2(t)$ and 
$\tilde{U}_{\mathrm{pos}}(t)$. The values of $\eta$ are in
agreement with previous measurements. Our results show
that MD simulations can be used for the determination
of critical exponents from the short-time behaviour. The dynamic critical
exponent changes from $z \approx 2$ for MC to $z \approx 1$ for MD
simulations.

The bond orientational order was studied with conventional MC dynamics.
The time evolution of the order parameter and the cumulant 
is not given by a simple power law behaviour, i.e.\  the scaling behaviour 
is not of the form (\ref{Eqscale}).
This phenomena could not be explained by a weak first-order
phase transition or by the way of determining neighbours.
A possible mixing of order parameter was discussed.

%%%%%%%%%%%%%%%%%%%%%%%%%%%%%%%%%%%%%%%%%%%%%%%%%%%%%%%%%%%%%%%%%%%%%%%%        
\section*{Acknowledgement}
Critical comments by  Lothar Sch\"{u}lke 
are gratefully acknowledged. Especially I benefitted from discussions with 
Conny Deiters. This work was supported in part
by the Deutsche Forschungsgemeinschaft under Grant No.\ DFG Schu 95/9-1.

\end{document}